\documentclass[epjST]{svjour}
\usepackage{dsfont}
\newcommand{\ud}{\mathrm{d}}

\usepackage{amsmath}
\usepackage{graphicx}
\usepackage{enumitem}

\usepackage{amssymb}
\usepackage{bclogo}
\usepackage{relsize}
\usepackage{dsfont}
\usepackage{bm}
\usepackage{graphbox}

\begin{document}

\title{Information geometry of scaling expansions of non-exponentially growing configuration spaces}

\author{Jan Korbel \inst{1,2} \fnmsep\thanks{\email{jan.korbel@meduniwien.ac.at}} \and Rudolf Hanel \inst{1,2} \fnmsep\thanks{\email{rudolf.hanel@meduniwien.ac.at}} \and Stefan Thurner \inst{1,2,3,4} \fnmsep\thanks{\email{stefan.thurner@meduniwien.ac.at}} }
\institute{Section for Science of Complex Systems, CeMSIIS, Medical University of Vienna, Spitalgasse 23, 1090 Vienna, Austria \and Complexity Science Hub Vienna, Josefst\"{a}dter Strasse 39, 1080 Vienna, Austria \and Santa Fe Institute, 1399 Hyde Park Road, Santa Fe, NM 87501, USA \and IIASA, Schlossplatz 1, 2361 Laxenburg, Austria}

%
%

\abstract{
Many stochastic complex systems are characterized by the fact that their configuration space doesn't grow exponentially as a function of the degrees of freedom.
The use of scaling expansions is a natural way to measure the asymptotic growth of the configuration space volume in terms of the scaling exponents of the system.
These scaling exponents can, in turn, be used to define universality classes that uniquely determine the statistics of a system.
Every system belongs to one of these classes.
Here we derive the information geometry of scaling expansions of sample spaces.
In particular, we present the deformed logarithms and the metric in a systematic and coherent way. We observe a phase transition for the curvature. The phase transition can be well measured by the characteristic { length $r$, corresponding to a ball with radius $2r$ having the same curvature as the statistical manifold}. Increasing characteristic length with respect to size of the system is associated with sub-exponential sample space growth is associated to strongly constrained and correlated complex systems. Decreasing of the characteristic lenght corresponds to super-exponential sample space growth that occurs for example in systems that develop structure as they evolve. Constant curvature means exponential sample space growth that is associated with multinomial statistics, and traditional Boltzmann-Gibbs, or Shannon statistics applies. This allows us to characterize transitions between statistical manifolds corresponding to different families of probability distributions.
}
%
%
%
%
%
\maketitle

\section{Introduction}

Statistical physics of complex systems has turned into an increasingly important topic with many applications.
Its main aim is to come up with a unified approach to understand, describe, and predict the statistical properties of a plethora of different complex systems; see e.g., \cite{Thurnerbook} for an overview.
While the microscopic nature of complex systems can be very different, their statistical properties often have common features across various systems.
Entropy is undoubtedly the key concept in statistical physics that connects the statistical description of microscopic dynamics with the macroscopic thermodynamic properties of a system.
The notion of entropy has been adopted also in other contexts, such as information theory or statistical inference, which are concepts quite different from thermodynamics  \cite{threefaces}.
One elegant and powerful concept arising from the theory of statistical inference is that of information geometry \cite{Aybook,Amaribook}.
It applies ideas from differential geometry to probability theory and statistics.
In this context, the concept of entropy also plays a crucial role, since the metric on the statistical manifold is derived from the corresponding (relative) entropy.
This, so-called Fisher-Rao metric, enables us to analyze statistical systems from a different perspective.
For example one can study critical transitions by calculating singularities of the metric \cite{Janke04}.

In information geometry, most attention has focused on systems that are governed by Shannon entropy \cite{Aybook,Amaribook}.
However, it is well known that many complex systems, especially strongly correlated or constrained systems, or systems with emergent components,
cannot be described within the framework of Shannon entropy \cite{Thurnerbook}.
For this reason, a number of generalizations to Shannon entropy have been proposed; in connection with power laws \cite{Tsallis88,Tsallis05}, special relativity \cite{Kaniadakis02}, multifractal thermodynamics \cite{Jizba04}, or black holes \cite{Tsallis13,Biro18}.

To classify entropies for { stochastic systems of various kinds}, it is natural to start with the information-theoretic foundations of Shannon entropy, i.e. the so-called
Shannon-Khinchin (SK) axioms \cite{Shannon48,Khinchinbook}. The first three SK axioms are usually formulated as:
\begin{itemize}
\item (SK1) Entropy is a continuous function of the probabilities $p_i$ only\footnote{ In several cases, entropies incorporate external parameter such as $q$ for Tsallis entropy or $c$ and $d$ for (c,d)-entropies. However, these parameters are constants that characterize the universality class of the process. They are not parameters subject to variation in entropy maximization.}.
\item (SK2) Entropy is maximal for the uniform distribution, $p_i=1/W$.
\item (SK3) Adding a state $W+1$ to a system with $p_{W+1}=0$ does not change the entropy of the system.	
\end{itemize}
%
The fourth axiom is called the \emph{composability axiom} and determines the entropy functional uniquely:
\begin{itemize}
\item (SK4) $H(A+B)=H(A)+H(B|A)$, where $H(B|A) = \sum p_k^{A} H(B|A_k)$
\end{itemize}
where $H(B|A_k)$ is the entropy of the conditional probability, $p_{B|A_k}$.
In this formulation, the unique solution that is compatible with SK1-4 is Shannon entropy $H(P) = - \sum_i p_i \log p_i$.
When the fourth axiom is relaxed, one can obtain wider class of entropic functionals.
{ First generalizations of the fourth axiom were introduced in connection with generalized additivity
\cite{abe2001,Ilic13} group laws \cite{Tempesta11} or statistical inference \cite{Jizba19}. These approaches are somewhat limited in scope, since they all lead to class of entropies, which can be expressed as a function of Tsallis entropy \cite{Tsallis88}}.

The relaxation of SK4 also naturally leads to a classification scheme of complex systems  \cite{ht11a,ht11b}.  The main idea of this approach is to study the asymptotic scaling exponents of the entropy functional that are associated to a particular system's configuration space. These systems are associated with systems that have a sub-exponentially growing configuration space, when seen as a function of degrees of freedom. { This classification scheme is based on \emph{mathematical} analysis of the asymptotic scaling of the entropic functionals that are governed by the first three SK axioms\footnote{ This does not mean that actual distribution functions that are, say, obtained from the maximum entropy principle must be equi-distributed, since the form of the distribution is determined not only by the entropic functional, but also by the constraints.}.}

Since the configuration space of most complex systems does not grow exponentially (as for the case of Shannon entropy),
but polynomially \cite{Tsallis05}, as a stretched exponential \cite{Anteneodo99}, or even super-exponentially \cite{jensen18},
the appropriate scaling behavior of the entropic functional is crucial for a proper thermodynamic interpretation.
For this end, we use recently developed \emph{scaling expansion} \cite{korbel18},
which is a special case of Poincar\'{e} asymptotic series \cite{Copsonbook}, whose coefficients are the scaling exponents of the system.

The aim of this paper is to define a generalization of Shannon entropy that matches the appropriate asymptotic scaling of a given system,
and use it to derive the associated generalized Fisher-Rao metric of the underlying statistical manifold.
For this end, we use the framework of deformed logarithms \cite{naudts02,naudts11}.
It has been shown recently \cite{korbel19} that one can naturally obtain two types of information metric within that framework,
one, corresponding to the maximum entropy principle with linear constraints, and the other, corresponding to the maximum entropy principle
when used with so-called escort constraints, instead of ordinary (linear) constraint terms.

Escort distributions appeared in connection with chaotic systems \cite{beck95},
and were discussed in the context of superstatistics \cite{beck03,tsallis03}.
Later it became possible to relate them to linear constraints through a \emph{log-duality} \cite{htg12}.
Interestingly, escort distributions  also appear as a canonical coordinate in information geometry \cite{abe03,ohara10}.
In this paper, we use both linear and escort approaches and compare their corresponding metric tensor and its invariants.
We focus particularly on the microcanonical ensemble in the thermodynamic limit,
since the metric should correspond to the system's asymptotic properties, given by its characteristic structure.
Some partial results for the curvature of escort metric were recently obtained in this direction \cite{ghikas18}.
However, no systematic and analytically expressible results for metric tensor and its scalar curvature have been obtained so far.
We show that the curvature of the statistical manifold naturally distinguishes between three types of systems:
systems with sub-exponentially growing configuration or sample space (correlated and constrained systems),
exponentially growing sample space (equivalent to ordinary multinomial statistics),
and super-exponentially growing sample space (e.g. systems that develop emergent structures as they evolve).
The vector of scaling exponents plays the role of a set of order parameters, i.e., the distance from the phase transition between sub-exponential and super-exponential phases.

The paper is organized as follows: Section 2 introduces the \emph{scaling expansion} and how to calculate corresponding scaling exponents.
We discuss several systems with non-trivial scaling exponents. In the last part of the section we establish a representation of
universality classes for complex systems, by introducing  \emph{scaling vectors} and their basic operations.
In Section 3, we briefly revisit the results of \emph{information geometry} in the framework of $\phi$-deformed logarithms.
We focus on information geometry with both, linear,  and escort constraints.
The main results of the paper are derived in Section 4, where we define the appropriate generalized logarithm by combining
the $\phi$-deformation framework and the requirement of asymptotic scaling.
The properties of corresponding entropic functionals are discussed. We exemplify the whole approach by the simple, yet very general, class of entropies with one correction term from the scaling expansion 
and calculate the asymptotic behavior of scalar curvature of the microcanonical ensemble in the thermodynamic limit.
The last section draws conclusions. The paper has several appendices that contain several technical details.

\section{Scaling expansion of the volume of configuration space}
The scaling expansion \cite{korbel18} is a method to investigate of the asymptotic scaling behavior of a sample space volume, $W(N)$.
Here $W$ is the number of accessible states in a system, and $N$ indicates size of a system \footnote{For example, think of $N$ as the number of particles in a system, or the number of throws in a coin tossing experiment.}.
The scaling expansion is a special case of the Poincar\'{e} asymptotic series, where the coefficients correspond to the scaling exponents of the system.
We introduce the notation for the iterated use of functions, $f^{(n)}(x) = \underbrace{f(\dots f(f(x))\dots)}_{n \ \mathrm{times}}$, to define a set of  re-scaling operations, $r_{\lambda}^{(n)}(x) = \exp^{(n)}(\lambda \log^{(n)}(x))$.
This set of re-scaling operations contains the well-known multiplicative re-scaling, $x \mapsto \lambda x$ ($n=0$),
power rescaling $x \mapsto x^\lambda$ ($n=1$), and the additive rescaling $x \mapsto x + \log \lambda$ ($n=-1$).
For each $n$, $r^{(n)}$ is a representation of the multiplicative group ($\mathds{R}^+$, $\times$), i.e., \mbox{$r^{(n)}_\lambda \circ r^{(n)}_{\lambda'} = r^{(n)}_{\lambda \lambda'}$}.
We now investigate how a function, $W(N)$, scales with re-scaling of $N \mapsto r^{(n)}_\lambda(N)$.
Note that due to a simple theorem (see Appendix A2 in \cite{korbel18}) the function $z(\lambda)$, defined as $z(\lambda)= \lim_{N \rightarrow \infty} \frac{g(r^{(n)}_\lambda(N))}{g(N)}$, must have the form $z(\lambda) = \lambda^c$ for $c \in \mathds{R}\cup\{\pm \infty \}$ whenever the limit exists.
We start with multiplicative scaling ($n=0$): The expression $\frac{W(\lambda N)}{W(N)}$ is, according to the theorem, equal to $\lambda^{c_0}$.
We assume that $W(N)$ is a strictly increasing function, then it follows that $c_0 \geq 0$\footnote{Details about processes with reducing sample space can be found e.g., in Refs. \cite{cht15,cht16,cth17,ht18}.}. It can happen that $c_0 = +\infty$.
In that case, the expression grows faster than any polynomial.
This problem can be resolved by using $\log^{(l)}(W(N))$ instead of $W(N)$, for an appropriate choice of $l$.
The parameter $l$ is chosen such that $c_0^{(l)}$, corresponding to $\frac{\log^{(l)} W(\lambda N)}{\log^{(l)} W(N)} \sim \lambda^{c_0^{(l)}}$, is finite.
We call $l$ the \emph{order} of the process.
We get that $W(N) \sim \exp^{(l)}(N^{c_0^{(l)}})$, for $N \gg 1$.
To get the corrections to the leading order, we use the fact that $\frac{\log^{(l)} W(\lambda N)}{\log^{(l)} W(N)} \frac{N^{c_0}}{(\lambda N)^{c_0}} \sim 1$.
When we use the re-scaling for $n=1$, we get the second scaling exponent: $\frac{\log^{(l)} W(N^\lambda)}{\log^{(l)} W(N)} \frac{N^{c_0^{(l)}}}{(N^\lambda)^{c_0^{(l)}}} \sim \lambda^{c_1^{(l)}}$.
Therefore, $W(N) \sim \exp^{(l)}(N^{c_0^{(l)}} (\log N)^{c_1^{(l)}})$.
One can continue along the same lines to obtain the asymptotic expansion of $W(N)$, which reads
\begin{equation}
W(N) \sim \exp^{(l)} \left(\prod_{j=0}^n (\log^{(j)} N)^{c_j^{(l)}} \right) \quad \ \mathrm{for} \ N \rightarrow \infty\, ,
\end{equation}
where $c_j^{(l)}$ are the characteristic scaling exponents. The scaling expansion of $\log^{(l)} W(N)$ can be written
\begin{equation}
\log (\log^{(l)} W(N)) = \sum_{j=0}^n c_j^{(l)} \log^{(j+1)} N + \mathcal{O}(\log^{n+1}(N)) \, .
\end{equation}
It can be shown that the scaling exponents can be calculated from $W(N)$ as
\begin{scriptsize}
\begin{equation}
c_k^{(l)} = \lim_{N \rightarrow \infty} \log^{(k)}(N)\left(\log^{(k-1)}(N)\left(\dots\left(\log(N)\left( N \, \frac{\ud \log^{(l)}(W(N))}{\ud N}- c_0^{(l)}\right)-c_1^{(l)}\right)\dots\right)-c_{k-1}^{(l)} \right).
\end{equation}
\end{scriptsize}

As a next step, we apply the scaling expansion to obtain the corresponding \emph{extensive} entropy functionals.
It is well-known that for complex systems (with sub- or super-exponential phase space growth) the Shannon-Boltzman-Gibbs entropy is not an extensive quantity.
To obtain an extensive expression for such systems, one can introduce an appropriate generalization of the entropy functional \cite{Thurnerbook}.
A natural way how to characterize thermodynamic entropy is to define the entropy functional $S(W)$ which is \emph{extensive}.
This requirement can be expressed for the microcanonical ensemble as $S(W(N)) \sim N$ for $N \rightarrow \infty$.
For the purpose of thermodynamics, we do not have to require exact extensivity (with equality sign), but only its weaker asymptotic version.
We consider the general trace-form entropy functional
\begin{equation}\label{eq:trf}
S(P) = \sum_{i=1}^W g(p_i).
\end{equation}
The scaling expansion of the extensive entropy in the microcanonical ensemble can be expressed as
\begin{equation}
 S(W) \sim \prod_{j=0}^{n} \log^{(j+l)}(W(N))^{d_j^{(l)}} \qquad \mathrm{for} \ N \rightarrow \infty \, ,
\end{equation}
and the scaling expansion of $g(x)$ is
\begin{equation}
g(x) \sim x \prod_{j=0}^n \log^{(j+l)}\left(\frac{1}{x}\right)^{d_j^{(l)}}.
\end{equation}
The scaling coefficients $d_j^{(l)}$ can be obtained by
\begin{scriptsize}
\begin{equation}
d_{k}^{(l)} = \lim_{N \rightarrow \infty} \log^{(l+k)}(W) \left(\log^{(l+k-1)}(W) \left(\dots\left(\log^{(l)}(W) \left(\left(N \, \frac{\ud \log^{(l)} W(N)}{\ud N}\right)^{-1}\right)-d_0^{(l)}\right)\dots\right)-d_{k-1}^{(l)}\right)\, .
\end{equation}
\end{scriptsize}
The requirement of extensivity determines the relation between scaling exponents $c_j^{(l)}$ and $d_j^{(l)}$ as
\begin{eqnarray}\label{eq:exponents}
d_0^{(l)} &=& \frac{1}{c_0^{(l)}}\nonumber\\
d_k^{(l)} &=& -\frac{c_k^{(l)}}{c_0^{(l)}} \quad k = 1,2,\dots \, .
\end{eqnarray}

\begin{figure}
\centering
\includegraphics[width=6.5cm]{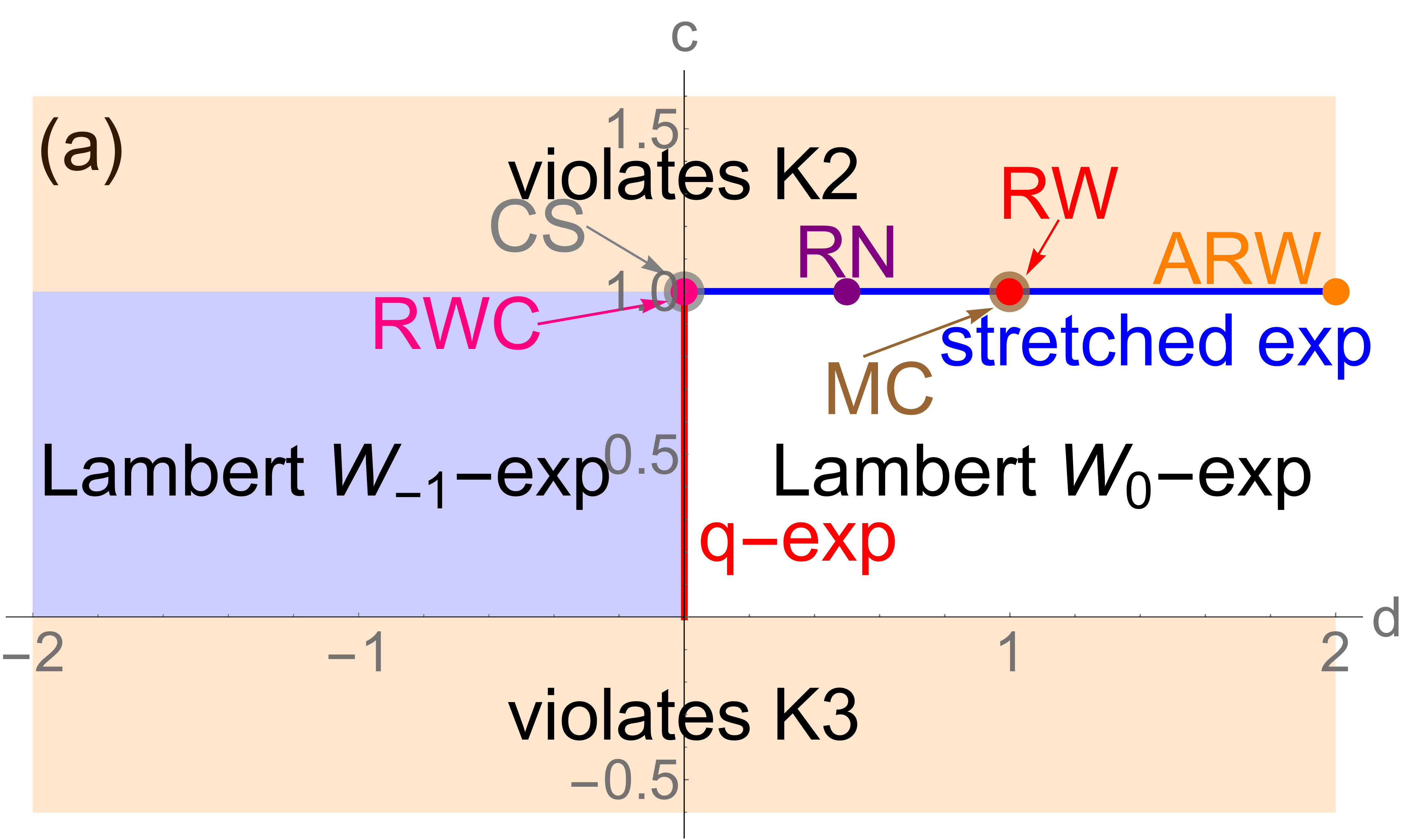}
\includegraphics[width=6cm]{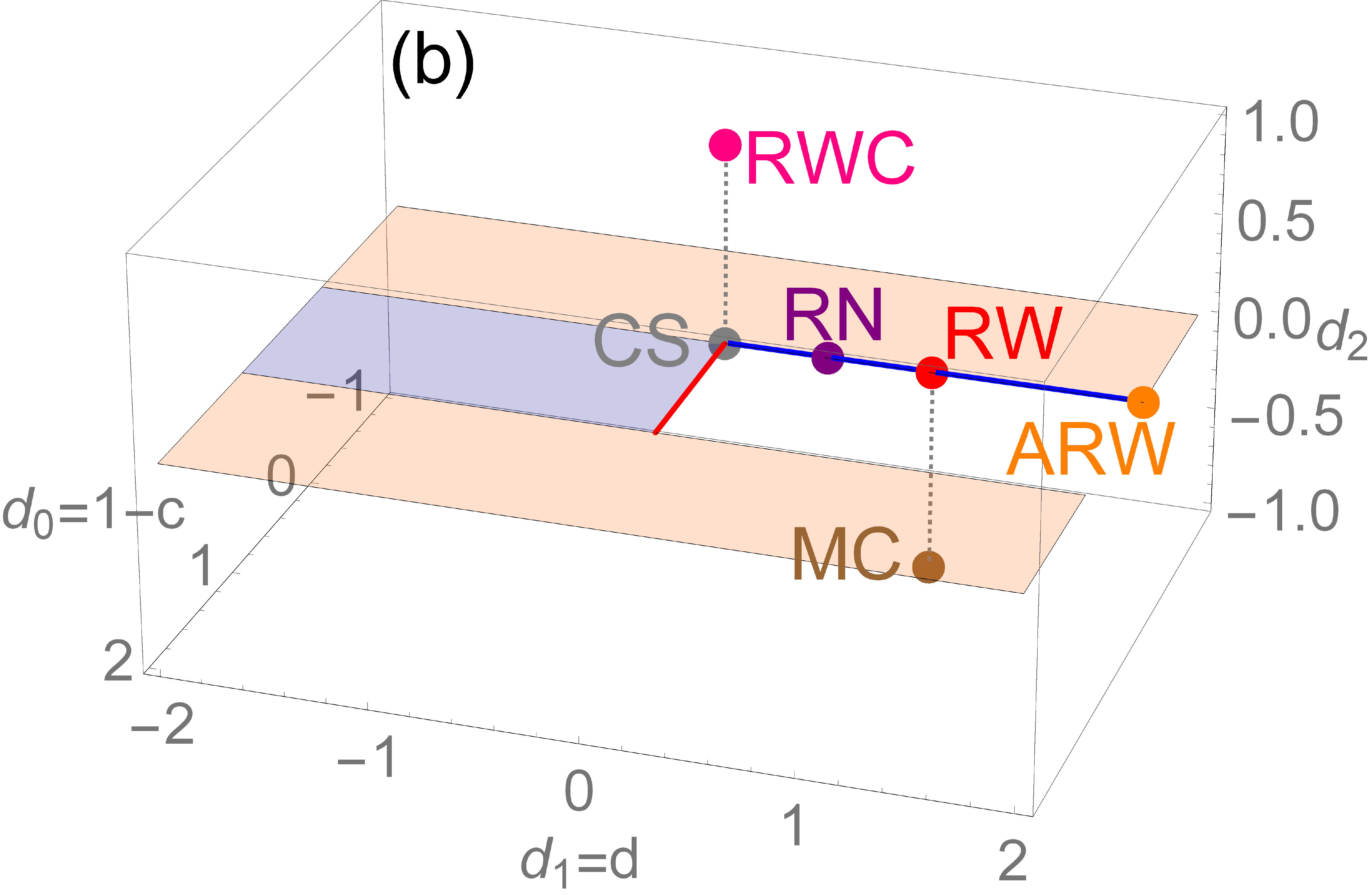}
\caption{parametric space of scaling expansion universality classes with scaling exponents of Random walk (RW), Aging random walk (ARW) Magnetic coin model (MC), Random networks (RN), Random walk cascade (RWC) and processes with compact support distribution (CS). (a) 2D parametric space of scaling expansion universality classes for the first two exponents (as in \cite{ht11a}).
We see that some super-exponential systems are not properly represented.
(b) Extension to three dimensions by adding the third scaling exponent, $d_2$.
All mentioned examples can be described with the first three scaling exponents.
}
\label{fig:exp}
\end{figure}

{\em Examples of systems with different scaling exponents.}
The first example is a random walk (RW) on the discrete one-dimensional lattice with two possible steps: left or right.
The space of all possible paths grows exponentially, $W_{RW}(N) =2^N \sim \exp(N)$, and we obtain the formula for Boltzmann entropy $S_{RW} = \log W_{RW}$
($k_B =1$).
Now consider an aging random walk (ARW) \cite{ht11b}, where the walker takes one step in a random direction, followed by two steps into a random direction, followed by three steps, etc.
In this case, the sample space grows sub-exponentially, $W_{ARW} \sim 2^{\sqrt{N}/2}$, and $S_{ARW} = (\log W_{ARW})^2$.
The next example is the magnetic coin model (MC) \cite{jensen18}, where each coin can be in two states: head or tail, however, two coins can also stick together and create a  bond state. It can be shown that the corresponding sample space grows super-exponentially, $W_{MC} \sim N^{N/2} e^{2 \sqrt N}$.
One can conclude that the corresponding extensive entropy is asymptotically equivalent to $S_{MC} = \log W_{MC}/\log \log W_{MC}$.
Another example of super-exponential processes are random networks (RN), whose sample spaces grow as $W_{RN} = 2^{\binom{N}{2}}$, and thus, $S_{RN} = (\log W_{RN})^{1/2}$.
The final example is the double-exponential growth of random walk cascade (RWC), where the walker can take a step to the right, to the left,
or split into two independent walkers \cite{korbel18}.
For this we get that $W_{RWC} = 2^{2^N}-1$, and, $S_{RWC} = \log \log W_{RWC}$.
In Fig. \ref{fig:exp} we show the parameter space of entropies given by three scaling exponents $(d_0,d_1,d_2)$. The above examples are indicated as points.
In Fig. \ref{fig:exp} (a)  the plane for the first two scaling exponents is shown, as presented in \cite{ht11a}.
We see that if one uses only the first two exponents, some super-exponential processes are not properly represented.
By adding a third scaling exponent this problem is solved Fig. \ref{fig:exp} (b).
So far, we have not yet found simple examples that need more than three scaling exponents.

\subsection{Universality classes for scaling expansions}
Scaling expansions define universality classes of statistical complex systems according to set of the scaling exponents of their sample space \cite{korbel18}.
The representation of the sample space volume, $W(N)$,  by its scaling expansion can be used to uniquely
describe the statistical properties in the thermodynamic limit.

Consider a function $c(x)$ represented by its scaling expansion
\begin{equation}
c(x) \sim \exp^{(l)}\left[\prod_{j=0}^n \left(\log^{(j)}(x)\right)^{c^{(l)}_{j}}\right] \; .
\end{equation}
Its scaling exponents can be collected in the \emph{scaling vector}
\begin{equation}\label{eq:scalingv}
\mathcal{C} =\{l;c^{(l)}_0,c^{(l)}_1,\dots,c^{(l)}_n\} \; .
\end{equation}
In principle, the scaling vector can be infinite, however, typically, after several terms the corrections are either zero, or do not contribute significantly.
The parameter $n$ denotes the number of corrections.

Let $a(x)$ and $b(x)$ be two functions with its respective scaling expansion determined by the two vectors of scaling exponents
\begin{equation}
\mathcal{A} = \{l_a;a_0,a_1,\dots,a_n\}
\end{equation}
\begin{equation}
\mathcal{B} = \{l_b;b_0,b_1,\dots,b_n\} \; .
\end{equation}
Without loss of generality, $n$ can be the same for both vectors because one can always append zeros to the shorter vector.
We can now define the {\em equivalence relation}
\begin{equation}
	a(x) \sim b(x) \ \mathrm{if}  \ \mathcal{A} \equiv \mathcal{B} \; ,
\end{equation}
as well as natural ordering
\begin{equation}
	a(x) \prec b(x) \ \mathrm{if}  \ \mathcal{A} < \mathcal{B} \; ,
\end{equation}
where the symbol $<$ is used in the lexicographic meaning, i.e.,
\begin{equation}\mathcal{A} < \mathcal{B} \qquad \mathrm{if} \quad \left\{
                                            \begin{array}{l}
                                              l_a < l_b\\
                                              l_a = l_b, a_0 < b_0\\
                                              l_a = l_b, a_0 = b_0, a_1 < b_1\\
                                                \dots \; .
\end{array}
                                          \right.
\end{equation}

For every vector $\mathcal{C}$ we define the corresponding \emph{entropy scaling vector} $\mathcal{D}$, denoted by $\mathcal{D} = \mathcal{C}^{-1}$, that is obtained from Eq. (\ref{eq:exponents}) by requirement of \emph{extensivity}. One can define analogous relations for $\mathcal{D}$ through the relations for corresponding vectors $\mathcal{C}$. Thus, for entropy scaling vectors $\mathcal{E}$ and $\mathcal{F}$, we can say that
\begin{equation}
\mathcal{E} < \mathcal{F} \qquad \mathrm{if} \quad \left\{
                                            \begin{array}{l}
                                              l_e < l_f\\
                                              l_e = l_f, e_0 < f_0\\
                                              l_e = l_f, e_0 = f_0, e_1 \mathbf{>} f_1\\
                                                \dots \; .
\end{array}
                                          \right.
\end{equation}
Note that for sub-leading scaling exponents the inequality is reversed, which is the result of Eq. (\ref{eq:exponents}).
Additionally, one can define basic algebraic operations on the scaling vectors, such as generalized addition or derivative operator.
More details can be found in Appendix \ref{scaling}. Let us make an important note. As discussed in \cite{korbel18}, the SK axioms set requirements on the admissible set of scaling exponents.
From SK2 we get that $d_l \equiv d_0^{(l)} >0$ and from SK3 that $d_0 < 1$.
Note that the vector $\mathcal{D}$ can be also represented as
\begin{equation}
\mathcal{D} =\{l;d^{(l)}_0,d^{(l)}_1,\dots,d^{(l)}_n\} = \{\underbrace{0,\dots,0}_{l \ \mathrm{times}},d_l,d_{l+1},\dots,d_{l+n}\} \; .
\end{equation}
This means that one can use the representation without specifying $l$ with an appropriate number of zeros at the beginning.
This is useful for example for the plots in the parametric space, where it is possible to plot processes of different order $l$
(as e.g., in Fig. \ref{fig:exp}).
However, one has to keep in mind that this representation can be misleading in the sense that the limit $d_l \rightarrow 0$ does not have a clear meaning,
since it changes the order of the process. This can be nicely seen in the example of Tsallis entropy \cite{Tsallis88}, where
\begin{equation}
\lim_{q \rightarrow 1^-} \frac{\sum_{i=1}^W p_i^q -1}{1-q} = - \sum_{i=1}^W p_i \log p_i \; ,
\end{equation}
which can be formulated in terms of entropy scaling vectors for as
\begin{equation}
 \lim_{q \rightarrow 1^-} \mathcal{D}  = \lim_{q \rightarrow 1^-} (1-q,0) = (0,1)
\end{equation}
Interestingly, the limit from above, $q \rightarrow 1^+$, is even more pathological.
In this case the scaling vector corresponding to $S_q(P)$ for $q>1$ is (0,0), because \mbox{$S_q(N) \sim N^{1-q}+1 \sim N^0$}.
These pathologies have their origin in the non-commutativity of limits,
$\lim_{N \rightarrow \infty} \lim_{d_l \rightarrow 0} \neq  \lim_{d_l \rightarrow 0} \lim_{N \rightarrow \infty}$.
The limit $d_l \rightarrow 0$ depends on the particular representation of the extensive entropy.
\\

\section{Information geometry of $\phi$-deformations}
Information geometry plays a central role in theory of information as well as in statistical inference.
It allows one to study the structure of the statistical manifold by means of differential geometry.
We derive the information-geometric properties of the scaling expansion in the framework of  $\phi$-deformed logarithms introduced in \cite{naudts02,naudts11}.
The $\phi$-deformation is a generalization of logarithmic functions.
It can be subsequently used to establish a connection with information theory,
where the logarithm plays the role of a natural information measure (Hartley information).
The $\phi$-deformed logarithm is defined by a positive, strictly increasing function $\phi(x)$, on $(0,+\infty)$ as
\begin{equation}
\log_\phi(x) = \int_{1}^x \frac{\ud y}{\phi(y)}\, .
\end{equation}
Hence, $\log_\phi$ is an increasing concave function with $\log_\phi(1)=0$. For $\phi(x) = x$, we obtain the ordinary logarithm.
Naturally,
\begin{equation}
\frac{\ud \log_\phi(x)}{\ud x} = \frac{1}{\phi(x)}\, .
\end{equation}
The inverse function of $\log_\phi$, the so-called $\phi$-exponential, is an increasing and convex function.
This enables one to define the parametric $\phi$-exponential family of probability distributions as
\begin{equation}
p(x;\theta) = \exp_\phi\left(\Psi(\theta) + \sum_i x_i \theta_i\right)\; ,
\end{equation}
where the function $\Psi(\theta)$ is called the \emph{Massieu function} and normalizes the distribution.
As discussed in \cite{korbel19}, there are two natural ways how to make a connection with the theory of information through the maximum entropy principle.
The first is based on the maximization of the entropy functional under the linear (thermodynamic) constraints,
the latter is based on a maximization under so-called escort (or geometric) constraints.
Both approaches lead to the $\phi$-exponential family.
The former approach defines the $\phi$-deformed entropy as \cite{naudts11}
\begin{equation}
S^N_\phi(p) = \sum_{i=1}^W \int_0^{p_i}\ud x \log_\phi(x)
\end{equation}
which is maximized by the $\phi$-exponential family for linear constraints, i.e., constraints of the type
\begin{equation}
\sum_{i=1}^W p_i E_i = \langle E \rangle.
\end{equation}
In information geometry, escort distributions play a special role of dual coordinates on statistical manifolds \cite{amari12}.
They can be defined by $\phi$-deformations as
\begin{equation}
P^{\phi}_i = \frac{\phi(p_i)}{\sum_k \phi(p_k)} = \frac{\phi(p_i)}{h_\phi(P)}\; .
\end{equation}
It can be shown that the entropy maximized by the $\phi$-exponential family for escort constraints, i.e., for constraints of the type
\begin{equation}
\sum_{i=1}^W P^\phi_i E_i = \langle E \rangle_\phi \; ,
\end{equation}
can be expressed as
\begin{equation}
S^A_\phi(p) = \sum_{i=1}^W P_i^\phi \log_\phi(p_i) = \frac{\sum_{i=1}^W \phi(p_i) \log_\phi(p_i)}{\sum_{j=1}^W \phi(p_j)}\; .
\end{equation}
For both approaches can be linked to information geometry, i.e., to derive a generalization of a Fisher information metric,
which can be done through a divergence (or relative entropy) of Bregmann type, which is defined as
\begin{equation}
D_f(p||q) = f(p)- f(q) - \langle \nabla f(q),p-q \rangle \; ,
\end{equation}
where $\langle \cdot,\cdot \rangle$ denotes the inner product. { Alternatively, one can use the divergence of Csisz\'{a}r type, but its information geometry is trivial, because it is conformal to ordinary Fisher information geometry, see e.g., Refs. \cite{korbel19,naudts02}}.

Let us consider a parametric family of distributions $p(\theta)$. The Fisher information metric of this family at point $\theta_0$ can be calculated as
\begin{equation}
g^f_{ij}(\theta) = \frac{\partial^2 D_f(p(\theta_0)||p(\theta))}{\partial \theta_i \partial \theta_j}|_{\theta=\theta_0}\; .
\end{equation}
Let us consider a discrete probability distribution $\{p_i\}_{i=0}^n$. The normalization is given by $\sum_{i=0}^n p_i =1$, so we consider $p_i,\dots,p_n$ as independent variables, while $p_0$ is determined from $p_0 = 1-\sum_{i=1}^n p_i$. { We parameterize this probability simplex by a $\phi$-deformed exponential family\footnote{Note that this parametric family typically constitutes a smooth manifold~\cite{amari12}.}}.  For the entropy $S_\phi^N$, we have $f_\phi^N(p) = \sum_i \int_0^{p_i} \log_\phi(x)\, \ud x$ while for $S_\phi^A(p)$ we end with $f(p) = \sum_i P_i^\theta \log_\phi(p_i)$. After a straightforward calculation, we obtain that \cite{korbel19}
\begin{equation}
g^{N}_{\phi,ij}(P) = \log_\phi'(p_i) \delta_{ij}+\log_\phi'(p_0) = \frac{1}{\phi(p_i)}\delta_{ij}+\frac{1}{\phi(p_0)}\; ,
\end{equation}
and
\begin{equation}
g^{A}_{\phi,ij}(P) =  - \frac{1}{h_\phi(p)} \left(\frac{\log_\phi''(p_i)}{\log_\phi'(p_i)} \delta_{ij}+ \frac{\log_\phi''(p_0)}{\log_\phi'(p_0)}\right) =  \frac{1}{h_\phi(p)}\left( \frac{\phi'(p_j)}{\phi(p_j)} \delta_{ij} + \frac{\phi'(p_0)}{\phi(p_0)}\right)\; ,
\end{equation}
respectively. As a result, for a given $\phi$-deformation there are two types of metric on the information manifold.
Note that it is natural to consider a one-parametric class of affine connections for which we obtain the so-called \emph{dually-flat structure} for which the corresponding Christoffel coefficients vanish \cite{amari12}.
This structure is useful in information geometry, however, we stick to the well-known Levi-Civita connection
(which can be obtained as a special case of a dually-flat connection, since the Levi-Civita connection is the only self-dual connection \cite{Amaribook}),
because the metric is non-vanishing.
Thus, the corresponding invariants, such as scalar curvature, are non-trivial and reveal some information about the statistical manifold.

Let us now focus on the scalar curvature of corresponding to the metric tensor, $R_\phi = g_\phi^{ik} g_\phi^{lj} R_{\phi,ilkj}$,
in the thermodynamic limit $N \rightarrow \infty$.
We focus on the microcanonical ensemble, i.e., we consider $p_i = 1/W$.
We assume no prior information about the system or its dynamics, so all states are equally probable.


It is possible to show in a technical but straightforward calculation that the scalar curvature can be expressed as { (see also \cite{zhang04,amari012})}
{
\begin{equation}
R_\phi(W) = \frac{W(W-1)}{(2 r_\phi(W+1))^2} \; ,
\end{equation}
}
which corresponds to the scalar curvature of a $W$-dimensional ball of radius $2r_\phi$.
The function $r_\phi$ depends only on the form of the $\phi$-deformation. We call the function $r_\phi$ { \emph{characteristic length}}.
For the case of the Amari metric, it can be expressed as
\begin{equation}\label{eq:ra}
(r^A_\phi(W))^2 = - \frac{\log_\phi'\left(\frac{1}{W}\right)^2 \log_\phi''\left(\frac{1}{W}\right)^3}{\left(\log_\phi'''\left(\frac{1}{W}\right) \log_\phi'\left(\frac{1}{W}\right)-3 \log_\phi''\left(\frac{1}{W}\right)^2\right)^2} \; ,
\end{equation}
while for the metric of Naudts type we obtain
\begin{equation}\label{eq:rn}
(r^N_\phi(W))^2 = \frac{W(\log_\phi'\left(\frac{1}{W}\right))^3}{(\log_\phi''\left(\frac{1}{W}\right))^2}\, .
\end{equation}

\section{Information geometry of scaling expansions}

Let us now consider an arbitrary $\phi$-deformed logarithm.
We show how to introduce a generalization of the logarithm with a given asymptotic scaling.
In contrast to $\phi$-deformations, we do not start with the definition of $\phi$, but focus on the definition of the logarithm.
We denote the desired logarithmic function as $\Lambda_\mathcal{D}$.
Let us state the requirements that $\Lambda_\mathcal{D}$ should fulfil:
\begin{enumerate}
\item \emph{Domain}: $\Lambda_{\mathcal{D}}: \mathds{R}^+ \rightarrow \mathds{R}$,
\item \emph{Monotonicity}: $\Lambda_{\mathcal{D}}'(x) > 0$,
\item \emph{Concavity}: $\Lambda_{\mathcal{D}}''(x) < 0$,
\item \emph{Normalization}: $\Lambda_{\mathcal{D}}'(1) = 1$,
\item \emph{Self-duality}: $\Lambda_{\mathcal{D}}(1/x) = - \Lambda_{\mathcal{D}}(x)$,
\item \emph{Scaling expansion}: $\Lambda_\mathcal{D}(x) \sim \prod_{j=0}^{k} \left[\log^{(j+l)}(x)\right]^{d_j^{(l)}}$ for $x \rightarrow \infty$.
\end{enumerate}
The requirements follow the properties of the ordinary logarithm.
Particularly convenient is the self-duality requirement, from which we can directly calculate the asymptotic expansion around $0^+$.
A direct consequence of self-duality is that
$\Lambda_{\mathcal{D}}(1) = 0$.
Next, we want to find a representation that is simple, analytically expressible, and universal for any set of scaling exponents.
Due to the self-duality requirement, we can focus only on the interval $(1,+\infty)$, while on the interval $(0,1)$ the logarithm is defined by the self-duality.
To find an appropriate representation, we start from the scaling expansion itself.
Unfortunately, the scaling expansion, $\prod_{j=0}^{k} \left[\log^{(j+l)}(x)\right]^{d_j^{(l)}}$, is not generally defined on the whole interval $(1,\infty)$,
since the domain of $\log^{(l)}(x)$ is $(\exp^{(l-2)}(1), \infty)$.
We can overcome this issue by adjusting the nested logarithm by replacing $\log \mapsto 1+\log$.
Further, to be able to fulfil the normalization condition, we add a multiplicative constant to the first nesting,
so that for each order the corresponding term can be expressed as $(1+r_j \log ([1+\log]^{(j-1)}(x))$.
Thus, the generalized logarithm can be expressed as
\begin{equation}
\Lambda_\mathcal{D}(x) = R \left[ \prod_{j=0}^n \left(1+r_j \log [1+\log]^{(j+l-1)}(x)\right)^{d_j^{(l)}} - 1\right] \; .
\end{equation}
The logarithm automatically fulfils the condition $\Lambda_\mathcal{D}(1)=0$.
The parameters $r_n$ define the set of scale parameters that influence the behavior at finite values,
while the asymptotic properties are preserved. Because
\begin{equation}
\Lambda'_\mathcal{D}(1) = R \sum_{j=0}^n r_j d_j^{(l)}
\end{equation}
we can obtain normalization of the derivative in several ways. For this we define the ``calibration''
\begin{eqnarray}\label{eq:cal}
  r_0 &=& \rho \frac{1- r \sum_{j=1}^n d_j^{(l)}}{r d_0^{(l)}} \\
  r_k &=& \rho\\
  R &=& r/\rho \; ,
\end{eqnarray}
where $r$ and $\rho$ are free parameters.
The parameter $\rho$ can be determined by additional requirements.
The first option is to require that $\Lambda_\mathcal{D}$ is smooth enough, at least it has continuous second derivative.
From the second derivative of the self-duality condition together with the normalization condition, we get
$\Lambda''_\mathcal{D}(1)= -1$.
Following a straightforward calculation, we find
\begin{equation}
\Lambda''_\mathcal{D}(1) = R\left( 2\sum_{i < j} r_i r_j d_i^{(l)} d_j^{(l)} + \sum_{j=0}^n \left(r_j^2 d_j^{(l)} \, . \left(d_j^{(l)}-1\right) - (j+l)r_j d_j^{(l)}\right) \right) \; .
\label{eq:cal2}
\end{equation}
Using Eq. (\ref{eq:cal}) in Eq. (\ref{eq:cal2}), we get an expression for $\rho_C$, i.e., the scale parameter in the smooth calibration
\begin{equation}
\rho_C = \frac{r \sum_{j} d_j^{(l)}+l-1}{\frac{d_0^{(l)}-1}{d_0^{(l)}r} \left(1- r \sum_{j} d_j^{(l)}\right)^2+  (2-r) \sum_{j} d_j^{(l)} - 2r \sum_{j\leq i} d_i^{(l)}d_j^{(l)} + r \sum_{j} \left(d_j^{(l)}\right)^2}\, .
\end{equation}
The free parameter $r$ can be used to ensure that $\rho$ is positive. Alternatively, we can simply consider $r_0=1$, which is useful for several applications. In this case we get that $\rho_L$ scale parameter in the leading-order calibration is simply
\begin{equation}
\label{eq:calib}
\rho_L = \frac{r d_0^{(l)}}{1-r \sum_{j=1}^n d_j^{(l)}} \; .
\end{equation}
Note that after a proper normalization, this calibration corresponds to the calibration used in \cite{ht11a,ht11b}.
Unless a continuous second derivative is explicitly required, it is more convenient to work with this simpler calibration.

\begin{figure}
\includegraphics[align=t,width=6.5cm]{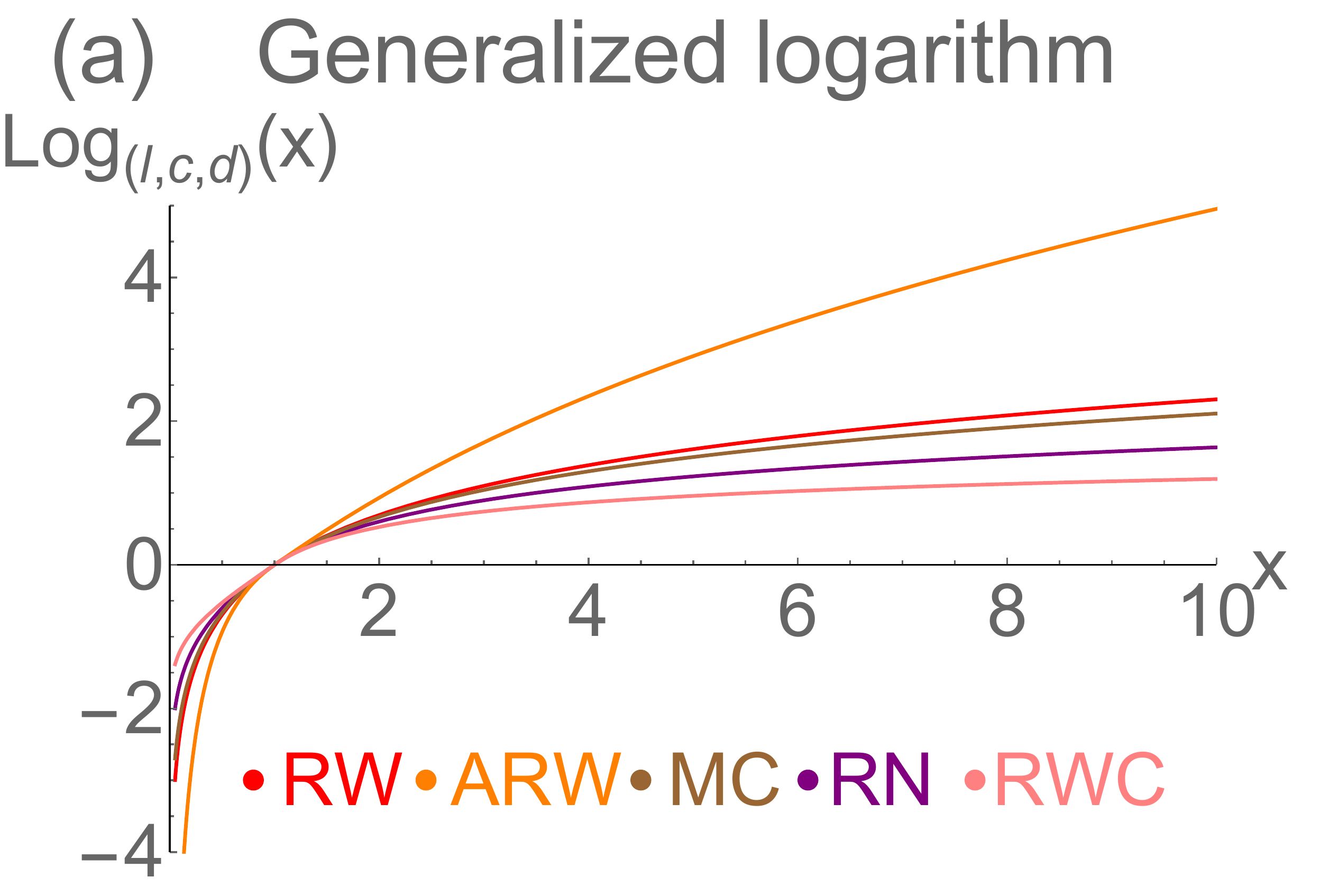}
\includegraphics[align=t,width=6.5cm]{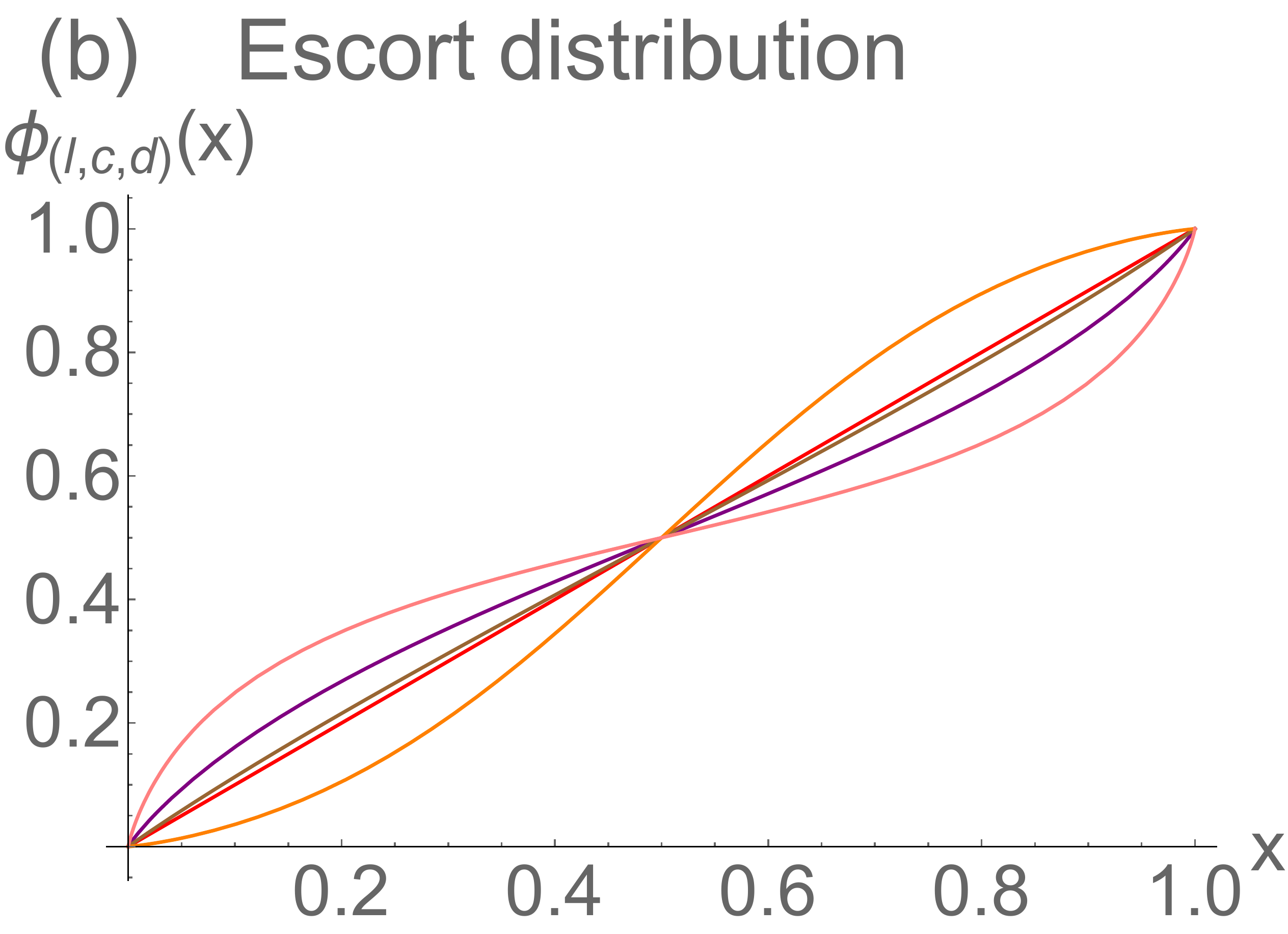}
\caption{(a) Generalized logarithms corresponding to scaling exponents of the aforementioned models. (b) Escort distributions corresponding to the generalized logarithms. The scaling exponents $(l;c,d)$ for the models are: Random walk (RW): $(1;1,0)$, Ageing random walk (ARW): $(1;2,0)$, Magnetic coin model (MC): $(1;1,-1)$, Random network (RN): $(1;1/2,0)$, Random walk cascade (RWC): $(2;1,0)$. }
\label{fig:log}
\end{figure}

\begin{figure}
\includegraphics[width=6.5cm]{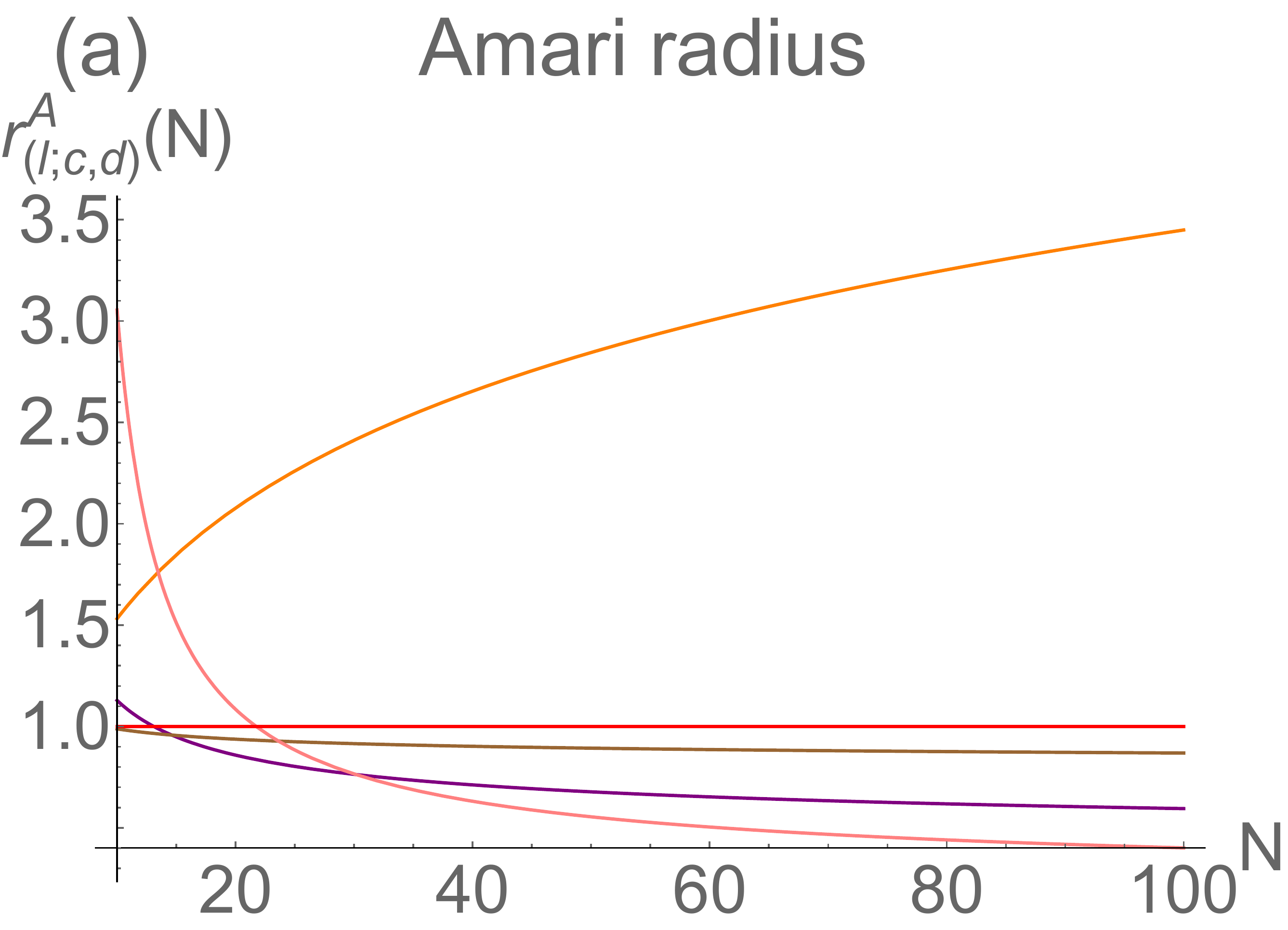}
\includegraphics[width=6.5cm]{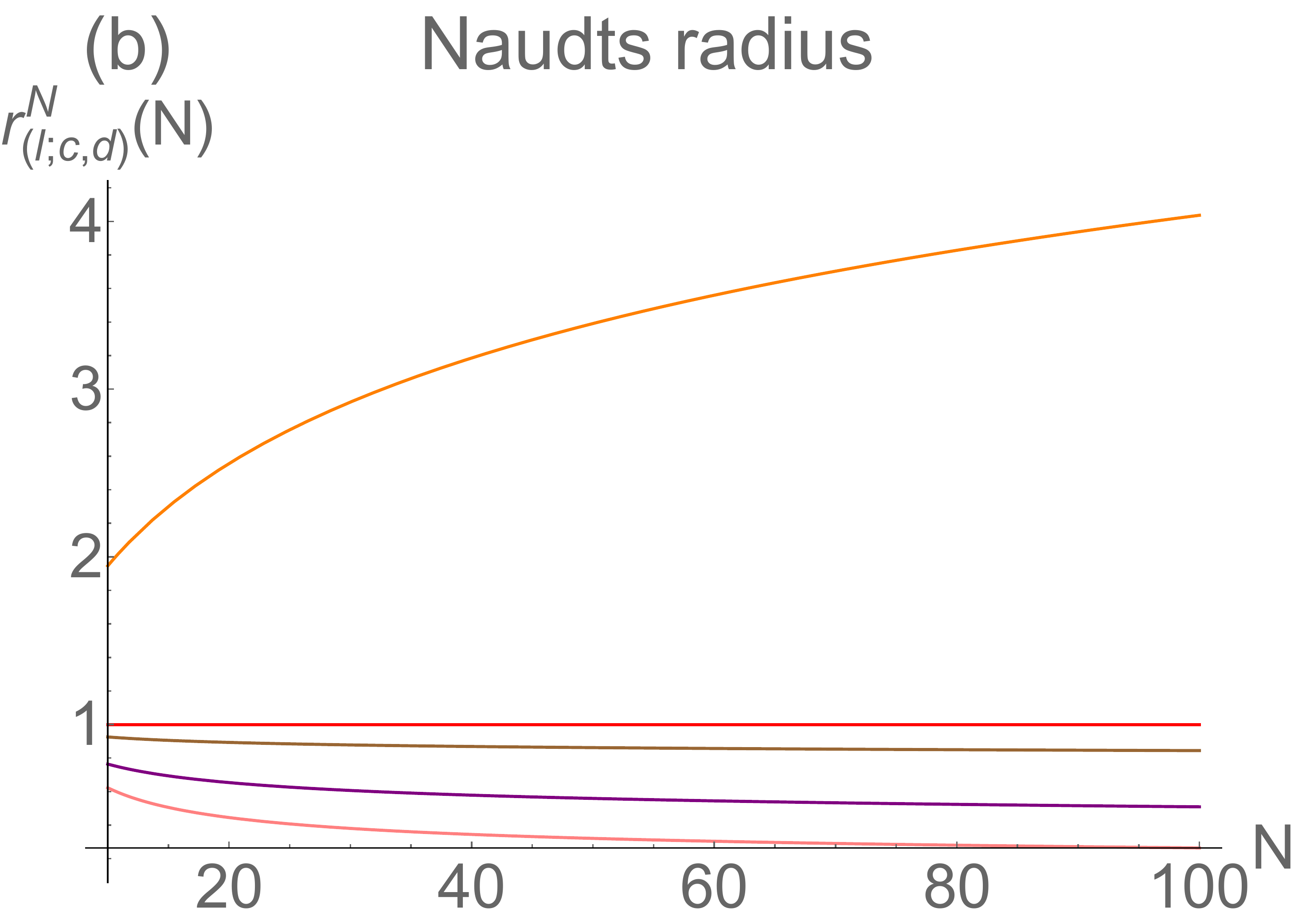}
\caption{Characteristic length corresponding to the curvature of the statistical manifold for equiprobable distribution corresponding to different scaling exponents. Figure (a) corresponds to length of Amari type, (b) corresponds to length of  Naudts type, respectively.
}
\label{fig:ig}
\end{figure}

We now turn our attention to the information geometry of $\Lambda_{\mathcal{D}}$-deformations and introduce a notation for the nested logarithm
\begin{equation}
\mu_k(x) = [1+\log]^{(k)}(x) \; .
\end{equation}
We sketch the results on for the scaling expansion with one correction.
All technical details can be found in Appendix \ref{lcd}.
In Appendix \ref{curv} we show the calculation for arbitrary scaling vectors and calibrations,
which is technically more difficult, but leads to the same results.
We now denote the scaling vector as $\mathcal{D} =(l;c,d)$.
Note that this entropy has been studied for $l=0$ in \cite{ht11a}.
This inspires us to define the generalized logarithm as
\begin{equation}\label{eq:lcd}
\log_{(l;c,d)}(x) = r \left(\mu_l(x)^c \left(1+\frac{1-cr}{dr} \log \mu_l(x)\right)^d-1\right) = \log_{(c,d)}\left(\mu_l(x)\right)\; .
\end{equation}
This definition corresponds to the choice of $\rho$ in Eq. (\ref{eq:calib})\footnote{Note that the original $(c,d)$-logarithm (as appearing in the rightmost part of Eq.~(\ref{eq:lcd})) was introduced in \cite{ht11a} for $l=0$ and $c \mapsto 1-c$.
Nevertheless, that choice of parametrization is not so convenient for $l > 0$.}.
The logarithms are depicted in Fig.~\ref{fig:log}(a) for various scaling exponents.
The inverse function, the deformed exponential, can be obtained in terms of the Lambert-W function\footnote{The
Lambert-W function  is defined as the solution of the equation $W(z)e^W(z) = z$.}
\begin{equation}
\exp_{(l;c,d)}(x) = \nu_{l}\left(\exp\left(-\frac{d}{c}\left[W\left(B(1-x/r)^{1/d}\right)-W(B) \right] \right)\right) \; ,
\end{equation}
where $B= \frac{cr}{1-cr}\exp\left(\frac{cr}{1-cr}\right)$ and $\nu_{l}$ is the inverse function of $\mu_{l}$, i.e.,
\begin{equation}
\nu_l(x) = \underbrace{\exp(\exp(\dots(\exp(x-1)-1)\dots))}_{l \ \mathrm{times}}\; .
\end{equation}
Note that depending on the values of $c$ and $d$ this deformed exponential contains the exponential, power laws, and stretched exponentials, respectively
\cite{Thurnerbook}.
It is easy to see that the corresponding scaling vector of the exponential is $\mathcal{C} = (l;1/c,-d/c)$.
The function $\phi_{(l;c,d)}(x)$ can be expressed as
\begin{eqnarray}
\phi_{(l;c,d)}(x) &=& \phi_{(0;c,d)}(\mu_l(x)) \cdot \mu_l(x)^{-1} \nonumber\\
&=& \frac{\mu_l(x)}{\log_{(c,d)-r}} \frac{d r + (1-c r) \log \mu_l(x)}{d+c\left(1-c r\right) \log \mu_l(x)} \prod_{j=0}^{l-1} \mu_k(x)\; .
\end{eqnarray}
The escort distribution, $\rho_{(l;c,d)}(p) =  \phi_{(l;c,d)}(p)/(\phi_{(l;c,d)}(p)+\phi_{(l;c,d)}(1-p))$,
corresponding to the two-event distribution $(p,1-p)$ is depicted in Fig.~\ref{fig:log}(b) for various scaling exponents.
Interestingly, for $\mathcal{D} < (1;1)$, i.e., for entropies corresponding to sub-exponential sample space growth, the distribution
shows high probabilities (generally $p >1/N$), while for $\mathcal{D} > (1;1)$, i.e.,
for super-exponential growth, the distribution shows low probabilities ($p < 1/N$).
Let us finally show the asymptotic behavior of the curvature that corresponds to the deformed logarithm.
It can be easily calculated if one keeps only dominant contributions from each term in the Eqs. (\ref{eq:ra}) and (\ref{eq:rn}).
In this case we have
\begin{eqnarray}
  \frac{\mathrm{d}^n \log_{(l;c)}(x)}{\mathrm{d} x^n}  \approx (\mu_l(x))^{c-1} \mu'_l(x) x^{1-n} \quad \mathrm{for} \ x \rightarrow \infty\; ,
\end{eqnarray}
and therefore
\begin{equation}
r_{(l;c)}(W) \approx 1/W \mu_l^{c-1}(W) \mu_l'(W) \quad  \mathrm{for} \ W \rightarrow \infty \; ,
\end{equation}
for both curvatures, calculated from both types of metric, as shown in \ref{lcd}.
From this we deduce that
\begin{equation}
\lim_{W \rightarrow \infty} r(W) = \left\{
  \begin{array}{ll}
    +\infty, & l=0 \ \mathrm{or} \ l=1, c>1 \\
    0, & l \geq 2 \ \mathrm{or} \ l=1, c<1 \; .
  \end{array}
\right.
\end{equation}
For the case $l=1$ and $c=1$, we can make a similar approximation
\begin{equation}
  \frac{\mathrm{d}^n \log_{(1;1,d)}(x)}{\mathrm{d} x^n}  \approx x^{-n} \log(1+\log(x))^d \; ,
\end{equation}
to get
\begin{equation}
r_{(1;1,d)}\approx \log(1+\log(W))^d \quad N \rightarrow \infty
\end{equation}
Similar results can be obtained for higher-order corrections.
The behavior of $r$ for different scaling vectors is depicted in Fig. \ref{fig:ig}.
We see that the asymptotic behavior is similar for both types of curvature, the only difference is for smaller $N$.

Similarly, we obtain the same behavior also for higher-order corrections (see Appendix \ref{curv}).
In conclusion, we find three distinct regimes for the statistical manifold with respect to the scaling vector

\begin{enumerate}[label=(\Roman*)]
\item $\mathcal{D} < (1;1)$ $\Leftrightarrow$ $r_{\mathcal{D}}(W) \rightarrow \infty$ for $W \rightarrow \infty$,
\item $\mathcal{D} = (1;1)$ $\Leftrightarrow$ $r_{\mathcal{D}}(W) = 1$ for $W > 0$,
\item $\mathcal{D} > (1;1)$ $\Leftrightarrow$ $r_{\mathcal{D}}(W) \rightarrow 0$ for $W \rightarrow \infty$.
\end{enumerate}

As a result, the curvature exhibits a phase transition --- the statistical manifold in thermodynamic limit is flattening for sub-exponential processes,
has constant section curvature for exponential processes,
and is curving for super-exponential processes.
While processes with exponentially growing sample space have (practically) independent sub-systems,
sub-exponential processes impose some restrictions and constaints on the sample space.
Super-exponential processes are characterized by emergent structures of its sample space.
The scaling vector plays the natural role of the set of order parameters. {Let us finally note that the limit $W \rightarrow \infty$ is performed for $r_\mathcal{D}(W)$. The ``limit space'' obtained in the limit of the statistical manifolds, for $W \rightarrow \infty$, might not be a smooth manifold and the curvature might not correspond to the limit $\lim_{W \rightarrow \infty} R_\mathcal{D}(W)$.}

\section{Conclusions and Perspectives}
In this paper, we have defined a class of deformed logarithms with a given scaling expansion in the framework of $\phi$-deformed logarithms.
The corresponding entropy can be used to define the statistical manifold with generalized Fisher-Rao metric.
We have shown that for the microcanonical ensemble in the thermodynamic limit, the scalar curvature exhibits a phase transition
where the critical point is represented by the class of phenomena that are characterized by exponentially growing phase spaces.
These include weakly interacting systems that are correctly described by Shannon entropy.
The scaling vector of a given system naturally defines a set of order parameters. A possible explanation for this phenomenon is that the number of independent degrees of freedom grows slower than the size of the system for sub-exponential processes and faster for the super-exponential processes. This classification, however, does not appear for the case of the Fisher metric of Csisz\'{a}r type, since the characteristic length is constant for every $\phi$-deformation.

Contrary to common approach in information geometry, where the statistical manifold corresponds to one functional family of distributions (e.g., exponential family of distributions), this paper presents a parametric way how to switch between different functional families of distributions (e.g., from power-laws to stretched exponentials). This opens a novel connection between parametric and non-parametric information geometry and enables to classify different types of statistical manifolds related to various classes of deformed exponential families.

It will be natural to extend these results to generalizations of { Bregmann divergence enabling gauge invariance \cite{naudts18}}. Moreover we will focus on application of the results to the canonical ensemble and use the well-known results using Fisher information metric on the thermodynamic manifold \cite{ruppeiner95,crooks97} for the case of complex systems, where we need to use the generalized form of the Boltzmann factor \cite{ht05}. Moreover, it should also be possible to go beyond equilibrium statistical mechanics and extend the generalized Fisher metric to non-equilibrium scenarios \cite{ito18}.








\section*{Acknowledgements}
We acknowledge support from the Austrian Science Fund FWF project I 3073.

\bibliographystyle{unsrt}

\newpage
\appendix

\section{Basic algebra of scaling vectors}
\label{scaling}
Let us discuss some definitions of ordinary operations on the space of scaling exponents. First, let us introduce a truncated vector of the scaling vector defined in Eq. (\ref{eq:scalingv}) as
\begin{equation}
\mathcal{C}_k =\{l;c^{(l)}_0,c^{(l)}_1,\dots,c^{(l)}_k\}
\end{equation}
where $k \leq n$. Then, we can introduce
\begin{itemize}
\item Truncated equivalence relation: $a(x) \sim^{(k)} b(x) \ \mathrm{if}  \ \mathcal{A}_k \equiv \mathcal{B}_k$
\item Truncated inequality relation: $a(x) \prec^{(k)} b(x) \ \mathrm{if}  \ \mathcal{A}_k < \mathcal{B}_k.$
\end{itemize}
Let us also add one set of inequality relations, and particularly for the case, when even the order $l$ is not equal. For this we define
\begin{itemize}
\item Strong inequality relation: $a(x) \ll b(x) \ \mathrm{if}  \ l_a < l_b\, .$
\end{itemize}

Let us  investigate representations of basic operations on the space of scaling exponents. Before that let us define the rescaling of the general operator $\mathcal{O}: \mathds{R}^m \mapsto \mathds{R}$ as
\begin{equation}
\mathcal{O}^{(l)}(x_1,x_2,\dots,x_m) = \exp^{(l)}\left[\mathcal{O}(\log^{(l)}x_1,\log^{(l)}x_2,\dots,\log^{(l)}x_m)\right].
\end{equation}
Let us now denote the generalized addition as $a(x) \oplus^{(l)} b(x)$ and multiplication as $a(x) \otimes^{(l)} b(x)$. It is easy to show that
\begin{equation}
a(x) \otimes^{(l)} b(x) = a(x) \oplus^{(l+1)} b(x)
\end{equation}
Let us now consider, without loss of generality, that $a(x) \prec b(x)$. The scaling vector $\mathcal{C}$ of $c(x) =  a(x) \otimes^{(l)} b(x)$
can be expressed as follows:
\begin{equation}
\mathcal{C} = \left\{
                \begin{array}{ll}
                   \mathcal{A} + \mathcal{B} = (l,a_0+b_0,a_1+b_1,\dots), & \mathrm{for} \ l_a = l_b = l; \\
                   \mathcal{B}, & \mathrm{for} \ l < l_a \leq l_b \ \mathrm{or} \ l = l_a < l_b ; \\
                  \mathrm{undefined}, & \mathrm{for} \ l > l_a.
                \end{array}
              \right.
\end{equation}
The scaling vector $\mathcal{C}$ of the generalized composition $c(x) = \exp^{(l)}b(\log^{(l)}a(x))$ can be expressed as
\begin{equation}
\mathcal{C} = \left\{
                \begin{array}{ll}
                   b_0^{(l_b)} \mathcal{A} = (l_a+l_b;a_0 b_0, a_1 b_0, a_2 b_0, \dots,a_n b_0), & \mathrm{for} \ l_a = l; \\
                   1^{(l_b)} \mathcal{A} = (l_a+l_b;a_0, a_1, a_2, \dots,a_n) , & \mathrm{for} \ l < l_a ; \\
                  \mathrm{undefined}, & \mathrm{for} \ l > l_a.
                \end{array}
              \right.
\end{equation}
Finally, let us focus on the derivative of the scaling expansion. Let us denote the rescaled derivative operator as
\begin{equation}
^{(l)}D_x[f] = \exp^{(l)}\left(\frac{\mathrm{d} (\log^{(l)} f(x))}{\mathrm{d} x}\right)
\end{equation}
The scaling vector corresponding to the rescaled derivative is
\begin{equation}
^{(l)}\mathcal{A}' = \left\{
                \begin{array}{ll}
                  (l_a;a_0-1,a_1,a_2,\dots,a_n), & \mathrm{for} \ l_a = l; \\
                   \mathcal{A},  & \mathrm{for} \ l_a > l ; \\
                  (l;\underbrace{-1,\dots,-1}_{l-l_a},0,\dots), & \mathrm{for} \ l_a < l.
                \end{array}
              \right.
\end{equation}
%
%
%
%

\section{Asymptotic curvature of $(l;c,d)$-logarithm}
\label{lcd}
In this appendix, we calculate asymptotic properties of the $(l;c,d)$ logarithm. Let us first express the derivatives of $(l;c,d)$ logarithm in terms of $(c,d)$ logarithm and $\mu_l$:
\begin{scriptsize}
\begin{eqnarray}
  \log'_{(l;c,d)}(x) &=& \log'_{(c,d)}(\mu_l(x)) \mu'_l(x)\, ,\\
  \log''_{(l;c,d)}(x)&=& \log''_{(c,d)}(x)(\mu'_l(x))^2 + \log'_{(c,d)}(\mu_l(x)) \mu''_l(x)\, , \\
  \log'''_{(l;c,d)}(x) &=& \log'''_{(c,d)}  (\mu'_l(x))^3 + 3 \log''_{(c,d)}(\mu_l(x)) \mu'_l(x) \mu''_l(x) + \log'_{(c,d)}(\mu_l(x)) \mu'''_l(x)\, .
\end{eqnarray}
\end{scriptsize}
The derivatives of the nested logarithm $\mu_l(x) = [1+\log]^{(l)}(x)$ can be expressed as:
\begin{scriptsize}
\begin{eqnarray}
\mu'_l(x) &=& \frac{1}{\prod_{k=0}^{l-1} \mu_k(x)}\, ,\\
\mu''_l(x) &=&  -\mu'_l(x) \sum_{k=0}^{l-1} \mu'_{k}(x) = -\frac{1}{\prod_{k=0}^{l-1} \mu_k(x)} \left(\sum_{k=0}^{l-1} \frac{1}{\prod_{j=0}^k \mu_j(x)}\right) \, ,\\
\mu'''_l(x) &=& -\mu''_l(x) \sum_{k=0}^{l-1} \mu'_{k}(x) - \mu'_l(x) \sum_{k=0}^{l-1} \mu''_{k}(x)\nonumber\\
&=&  \mu'_l(x) \sum_{j=0}^{l-1} \mu'_{j}(x) \sum_{k=0}^{l-1} \mu'_{k}(x) + \mu'_l(x)  \sum_{k=0}^{l-1}\mu'_k(x) \sum_{j=0}^{k-1} \mu'_{j}(x)\nonumber\\
&=& 2 \mu'_l(x)  \sum_{k=0}^{l-1}\mu'_k(x) \sum_{j=0}^{k-1} \mu'_{j}(x)+ \mu'_l(x)  \sum_{k=0}^{l-1}\mu'_k(x) \sum_{j=k}^{l-1} \mu'_{j}(x)\, .
\end{eqnarray}
\end{scriptsize}
Let us first denote $l_{(c,d)}(x) = \log_{(c,d)}(x)+r$. Then the derivatives of $\log_{(c,d)}$ can be expressed as (see also Ref. \cite{korbel19}):
\begin{scriptsize}
\begin{eqnarray}
  \log'_{(c,d)}(x) &=& \frac{l_{c,d}(x)}{{x (d r + (1-c r) \log x)}}\,\left[d+c\left(1-c r\right) \log x\right]\, , \\
\log''_{(c,d)}(x) &=& \frac{l_{c,d}(x)}{x^2 (d r + (1-c r) \log x)^2}\, \left[d \left(d-dr-(c r-1)^2\right)\right.\nonumber\\
&+& \left. d \left(c^2 (r-2) r+2c-1\right) \log x+(c-1) c (c r-1)^2 \log ^2 x\right]\, ,\\
\log'''_{(c,d)}(x) &=& \frac{ l_{c,d}(x)}{{x^3 (d r+(1-c r) \log x)^3}} \left[d \left(3 d (r-1) (c r-1)^2-2 (c r-1)^3+d^2 \left(2 r^2-3 r+1\right)\right)\right.\nonumber\\
&&+ d (c r-1)  \left(3 c^3 r^2-3 c^2 r (r+2)+c \left(d \left(-2 r^2+6 r-3\right)+6 r+3\right)+d (3-4 r)-3\right)\log x\nonumber\\
&&- d \left(3 c^2 (r-1)+c (6-4 r)-2\right) (c r-1)^2 \log ^2x\nonumber\\
&&- \left.c \left(c^2-3 c+2\right) (c r-1)^3 \log ^3 x\right]\, .
\end{eqnarray}
\end{scriptsize}
In the asymptotic limit, only dominant contributions are relevant. Thus, let us consider only dominant scaling $c$ (i.e., take $d=0$) and we get
\begin{scriptsize}
\begin{eqnarray}
  \log'_{(l;c,d)}(x) &=& \log'_{(c,d)}(\mu_l(x)) \mu'_l(x) \approx (\mu_l(x))^{c-1} \mu'_l(x)\, ,\\
  \log''_{(l;c,d)}(x)&\approx& \log'_{(c,d)}(\mu_l(x)) \mu''_l(x) \approx -\frac{(\mu_l(x))^{c-1} \mu'_l(x)}{x} \, \\
  \log'''_{(l;c,d)}(x) &\approx& \log'_{(c,d)}(\mu_l(x)) \mu'''_l(x) \approx \frac{(\mu_l(x))^{c-1} \mu'_l(x)}{x^2}\, .
\end{eqnarray}
\end{scriptsize}
Thus, we plugged in Eqs. (\ref{eq:ra},\ref{eq:rn}) then we get
\begin{scriptsize}
\begin{equation}
r_{(l;c)}^A \approx \frac{\left(\frac{\mu_l^{c-1}(x) (\mu_l'(x))}{x}\right) \left(\mu_l^{c-1}(x) (\mu_l'(x))\right)^2}{\left[\left(\mu_l^{c-1}(x) (\mu_l'(x))\right)\left(\frac{\mu_l^{c-1}(x) (\mu_l'(x))}{x^2}\right) - 3 \left(\frac{\mu_l^{c-1}(x) (\mu_l'(x))}{x}\right)^2    \right]^2} \approx x \mu_l^{c-1}(x) \mu_l'(x)\, ,
\end{equation}
\end{scriptsize}
and
\begin{scriptsize}
\begin{equation}
r_{(l;c)}^N \approx \frac{\left(\mu_l^{c-1}(x) (\mu_l'(x))\right)^3}{x \left(\frac{\mu_l^{c-1}(x) (\mu_l'(x))}{x}\right)^2} \approx x \mu_l^{c-1}(x) \mu_l'(x)\, .
\end{equation}
\end{scriptsize}
Let us then focus on the situation $l=1$, $c=1$. In this case the leading order terms cancel and we have to look at the first correction given by the scaling exponent $d$. In this case
\begin{scriptsize}
\begin{eqnarray}
  \log'_{(1;1,d)}(x) &\approx& \frac{\log(1+\log(x))^d}{x}\, , \\
  \log''_{(1;1,d)}(x)&\approx& \frac{\log(1+\log(x))^d}{x^2}\, ,\\
  \log'''_{(1;1,d)}(x) &\approx& \frac{\log(1+\log(x))^d}{x^3}\, .
\end{eqnarray}
\end{scriptsize}
So the curvature of both Amari and Naudts type can be asymptotically expressed as
\begin{scriptsize}
\begin{equation}
r_{(1;1,d)}\approx \log(1+\log(x))^d\, .
\end{equation}
\end{scriptsize}

\section{Fisher metric and scalar curvature corresponding to general logarithm}
\label{curv}
Let us now show the full calculation of the scalar curvature corresponding to the $\Lambda_\mathcal{D}$-logarithm with arbitrary scaling vector $\mathcal{D}$ and constants $r_j$. Let us first recall the product rule for higher derivatives. The first three derivatives of a function $\Lambda_\mathcal{D}(x) =  R \left(\prod_{j=0}^n \lambda_j(x) - 1\right)$:
\begin{scriptsize}
\begin{eqnarray}
\Lambda'_\mathcal{D}(x) &=&  R \left(\prod_{j=0}^{n} \lambda_j(x) \right) \left[\sum_{j=0}^n \frac{\lambda'_j(x)}{\lambda_j(x)}\right]\, ,\\
\Lambda''_\mathcal{D}(x) &=& R \left(\prod_{j=0}^{n} \lambda_j(x) \right) \left[2 \sum_{i < j} \frac{\lambda'_i(x)\lambda'_j(x)}{\lambda_i(x)\lambda_j(x)} + \sum_{j=0}^n \frac{\lambda''_j(x)}{\lambda_j(x)} \right]\, ,\\
\Lambda'''_\mathcal{D}(x) &=& R \left(\prod_{j=0}^{n} \lambda_j(x) \right) \left[6 \sum_{i < j < k} \frac{\lambda'_i(x)\lambda'_j(x)\lambda'_k(x)}{\lambda_i(x)\lambda_j(x)\lambda_k(x)}\right. \nonumber\\ &&\left. + 3 \sum_{i < j} \frac{\lambda''_i(x)\lambda'_j(x)+\lambda'_i(x)\lambda''_j(x)}{\lambda_i(x)\lambda_j(x)} + \sum_i \frac{\lambda'''_i(x)}{\lambda_i(x)} \right]\, .
\end{eqnarray}
\end{scriptsize}
The derivatives of $\lambda_j$ can be expressed by defining function $\mathcal{L}$
\begin{scriptsize}
\begin{equation}
\mathcal{L}_j(x) = \frac{1}{\left(1+r_j \log \mu_{j+l-1}(x)\right) \prod_{k=0}^{j+l-1} \mu_k(x)} = \frac{\mu'_{j+l}(x)}{\left(1+r_j \log \mu_{j+l-1}(x)\right)}\, .
\end{equation}
\end{scriptsize}
Then we can express
\begin{scriptsize}
\begin{eqnarray}
\lambda_j'(x) &=& \lambda_j(x) \left(r_j d_{j} \mathcal{L}_j(x)\right)\, ,\\
\lambda_j''(x) &=&  \lambda_j(x)  \left(r_j^2 d_{j}^2 \mathcal{L}^2_j(x) + r_j d_{j}\mathcal{L}'_j(x)\right)\, ,\\
\lambda_j'''(x) &=& \lambda_j(x) \left(r_j^3 d_j^3 \mathcal{L}^3_j(x) + 3 r_j^2 d_j^2 \mathcal{L}'_j(x) \mathcal{L}_j(x) + r_j d_j \mathcal{L}''_j(x) \right)\, .
\end{eqnarray}
\end{scriptsize}
The derivatives of $\mathcal{L}_j(x)$ can be expressed as
\begin{scriptsize}
\begin{eqnarray}
\mathcal{L}'_j(x) &=& -\mathcal{L}^2_j(x) \left(r_j  + (1+r_j \log \mu_{j+l-1}(x))\sum_{k=0}^{l+j-1} \prod_{m=k+1}^{j+l-1}\mu_m(x) \right)\, ,\\
\mathcal{L}''_j(x) &=& - 2 \mathcal{L}^3_j(x) \left(r_j  + (1+r_j \log \mu_{j+l-1}(x))\sum_{k=0}^{l+j-1} \prod_{m=k+1}^{j+l-1}\mu_m(x) \right)^2\nonumber \\
&& - \ \mathcal{L}^2_j(x) \left(r_j \sum_{k=0}^{l+j-1} \prod_{m=k+1}^{j+l-1}\mu_m(x) \right.\nonumber\\
&& \left. + \ (1+r_j \log \mu_{j+l-1}(x)) \sum_{k=0}^{j+l-1} \sum_{m=k+1}^{j+l-1} \frac{\prod_{p=k+1}^{j+l-1}\mu_p(x)}{\prod_{p'=0}^{m}\mu_{p'}(x)}  \right)\, .
\end{eqnarray}
\end{scriptsize}
We can finally rewrite the derivatives of $\Lambda_D$ as
\begin{scriptsize}
\begin{eqnarray}
\frac{\ud}{\ud x} \left(\Lambda_\mathcal{D}(x) \right) &=&  R \left(\prod_{j=0}^{n} \lambda_j(x) \right) \left[\sum_{j=0}^n \mathcal{C}^1_j(x) \mathcal{L}_j(x) \right]\, ,\\
\frac{\ud^2}{\ud x^2} \left(\Lambda_\mathcal{D}(x) \right) &=& R \left(\prod_{j=0}^{n} \lambda_j(x) \right) \left[\sum_{i=1}^n \sum_{j=1}^n \mathcal{C}^2_{ij}(x) \mathcal{L}_i(x)\mathcal{L}_j(x)\right]\, ,\\
\frac{\ud^3}{\ud x^3} \left(\Lambda_\mathcal{D}(x) \right) &=& R \left(\prod_{j=0}^{n} \lambda_j(x) \right) \left[\sum_{i=1}^n \sum_{j=1}^n \sum_{k=1}^n \mathcal{C}^3_{ijk}(x) \mathcal{L}_i(x)\mathcal{L}_j(x)\mathcal{L}_k(x)\right]\, ,
\end{eqnarray}
\end{scriptsize}
where the coefficients $\mathcal{C}$ can be expressed as
\begin{scriptsize}
\begin{eqnarray}
\mathcal{C}^1_i(x) &=& r_i d_i\, ,\\
\mathcal{C}^2_{ij}(x) &=&   r_i d_i \left[r_j d_j - \delta_{ij}\mathcal{A}_j(x)\right]\, ,\\
\mathcal{C}^3_{ijk}(x) &=& r_i d_i \left[r_j d_j r_k d_k - \left( \delta_{ij} r_j d_j \mathcal{A}_j(x) + \delta_{ik} r_k d_k \mathcal{A}_k(x) + \delta_{jk} r_j d_j \mathcal{A}_j(x) \right) - \delta_{ijk} \mathcal{B}_i(x) \right]\, ,
\end{eqnarray}
\begin{scriptsize}
where
\end{scriptsize}
\begin{eqnarray}
\mathcal{A}_i(x) &=& \left((r_i+ (1+r_i \log \mu_{i+l-1}(x)) )\sum_{k=0}^{i+l-1} \prod_{m=k+1}^{i+l-1}\mu_m(x)  \right)\, ,\\
\mathcal{B}_i(x) &=& 2 \mathcal{A}_i(x)^2 + \mathcal{L}_i(x) \left(r_i \sum_{k=0}^{l+j-1} \prod_{m=k+1}^{i+l-1}\mu_m(x)\right.\nonumber\\&& \left.+ (1+r_i \log \mu_{i+l-1}(x)) \sum_{k=0}^{i+l-1} \sum_{m=k+1}^{i+l-1} \frac{\prod_{p=k+1}^{i+l-1}\mu_p(x)}{\prod_{p'=0}^{m}\mu_{p'}(x)}  \right)\, .
\end{eqnarray}
\end{scriptsize}
Finally, we plug the expressions for $\Lambda_{\mathcal{D}}$ and its derivatives into Eqs. (\ref{eq:ra},\ref{eq:rn}), and we end with
\begin{scriptsize}
\begin{equation}
r^A_{\mathcal{D}}(x) = \frac{\left(\sum_i \mathcal{C}_i^1 \mathcal{L}_i(x)\right)^2 \left(\sum_{kl} \mathcal{C}^2_{kl}(x) \mathcal{L}_k(x) \mathcal{L}_l(x) \right)^3}{\left(\sum_{ijkl} \left(\left[\mathcal{C}_i^1(x) \mathcal{C}^3_{jkl}(x) -3 \mathcal{C}^2_{ij}(x) \mathcal{C}^2_{kl}(x)\right] \mathcal{L}_i(x) \mathcal{L}_j(x) \mathcal{L}_k(x) \mathcal{L}_l(x) \right)\right)^2}\, ,
\end{equation}
\end{scriptsize}
and
\begin{scriptsize}
\begin{equation}
r^N_{\mathcal{D}}(x) = \frac{\left(\sum_i C^1_i(x) \mathcal{L}_i(x)\right)^3}{x\left(\sum_{ij} C_{ij}^2(x)\mathcal{L}_i(x)\mathcal{L}_j(x)\right)^2}\, ,
\end{equation}
\end{scriptsize}
for $x = 1/W$, respectively.
%





\end{document}